\documentclass[twocolumn]{natureFIG}
\usepackage[dvips]{graphicx}
\usepackage{amssymb}
\usepackage{amsmath}
\usepackage{color}
\usepackage{xspace}
\newcommand{\ket}[1]{\ensuremath{|#1\rangle}\xspace}
\newcommand{\bra}[1]{\ensuremath{\langle #1|}\xspace}

\title{Coherent Electron-Phonon Coupling in Tailored Quantum Systems}

\author{P. Roulleau, S. Baer, T. Choi, F. Molitor, J. G\"uttinger, T. M\"uller, S. Dr\"oscher, K. Ensslin and  T. Ihn}
\date{\today}

\begin{document}

\maketitle

\begin{affiliations}
\item Solid State Physics Laboratory, ETH Zurich, 8093
Zurich, Switzerland
\end{affiliations}
\begin{addendum}
\item[Correspondence:] Correspondence and requests for materials
should be addressed to P.R.~(email: roulleau@phys.ethz.ch).
\end{addendum}
\begin{abstract}

The coupling between a two-level system and its environment leads to
decoherence. Within the context of coherent manipulation of
electronic or quasiparticle states in nanostructures, it is crucial
to understand the sources of decoherence. Here, we study the effect
of electron-phonon coupling in a graphene and an InAs nanowire
double quantum dot. Our measurements reveal oscillations of the
double quantum dot current periodic in energy detuning between the
two levels. These periodic peaks are more pronounced in the nanowire
than in graphene, and disappear when the temperature is increased.
We attribute the oscillations to an interference effect between two
alternative inelastic decay paths involving acoustic phonons present
in these materials. This interpretation predicts the oscillations to
wash out when temperature is increased, as observed experimentally.

\end{abstract}

Coherent spin manipulation has already been accomplished in
AlGaAs/GaAs double quantum dots
(DQDs)\cite{Petta30092005,Kopp06Nat442p766} and, more recently, also
in InAs nanowires (NWs) \cite{Nadj10Nat5p327}. While the coherence
times are usually limited by random nuclear fields
\cite{Bluhm10Nat10p1038}, also electron-phonon coupling can be a
source of decoherence \cite{Haya03PRL91p226804}. InAs nanowires (NW)
and graphene are two alternative and promising materials for
achieving coherent spin manipulation. In InAs NW DQDs, spin-orbit
interactions (SOI) are very strong and enable a more efficient
electron spin resonance driven by SOI compared to AlGaAs/GaAs DQDs
\cite{Nadj10Nat5p327}. In graphene, it is expected that hyperfine
coupling as a source of decoherence is very weak compared to
AlGaAs/GaAs. While electron-phonon interaction effects have been
observed in carbon nanotube
\cite{Leturcq09NP5p327,Escott10Nano21p274018}, AlGaAs/GaAs
\cite{Weig04PRL92p046804}, or silicon quantum dots (QDs)
\cite{Zwan09NL9p1071} and in AlGaAs/GaAs DQDs
\cite{Naber06PRL96p136807,Fuji98Science282p932}, only little is
known about electron-phonon interaction in graphene and InAs
nanowires.

Almost 60 years ago, Dicke predicted superradiant and subradiant
spontaneous emission \cite{Dicke54PR93p99}, which was observed 40
years later with two trapped ions \cite{DeVoe96PRL76p2049}. In this
experiment, the spontaneous emission rate $\Gamma(R)$ of a two-ion
crystal excited by a short laser pulse was studied as a function of
the ion-ion separation $R$. Superradiant (subradiant) spontaneous
emission was observed with $\Gamma(R)>\Gamma_0$
($\Gamma(R)<\Gamma_0$), where $\Gamma_0$ is the emission rate of a
single ion. In analogy to the Dicke subradiance phenomenon, Brandes
\textit{et al.} \cite{Brand99PRL83p3021} later proposed an
interference effect due to electron-phonon interactions in a
solid-state two-level system (DQD). Our experimental observations
are interpreted in this framework.

Here we report on an effect associated with coherent electron-phonon
coupling in two entirely different DQD systems and therefore
different electronic and phononic environments. The very strong
confinement of electronic states in these two materials, in contrast
to AlGaAs/GaAs DQDs, has enabled us to observe this coherent
coupling, the solid-state analogue of the Dicke subradiance
phenomenon. Our measurements show periodic oscillations in the
current through both double dot systems as a function of the energy
difference of the levels in the two dots. The energy dependence of
these oscillations allows us to infer a coherent coupling between
electrons and the phonon field. We find an enhancement of the
coherent oscillations in the InAs NW compared to graphene in
agreement with dimensional considerations. The temperature at which
the experimentally detected oscillations disappear ($\approx$ 600
mK) clearly supports the relevance of a coherent effect in the
coupled electron-phonon system. Finally, this study shows new
possibilities for using graphene and InAs nanowires as
nanoelectromechanical devices and, more specifically, as phonon
detectors.

\begin{figure}
\centerline{\includegraphics[angle=0,width=9cm,keepaspectratio,clip]{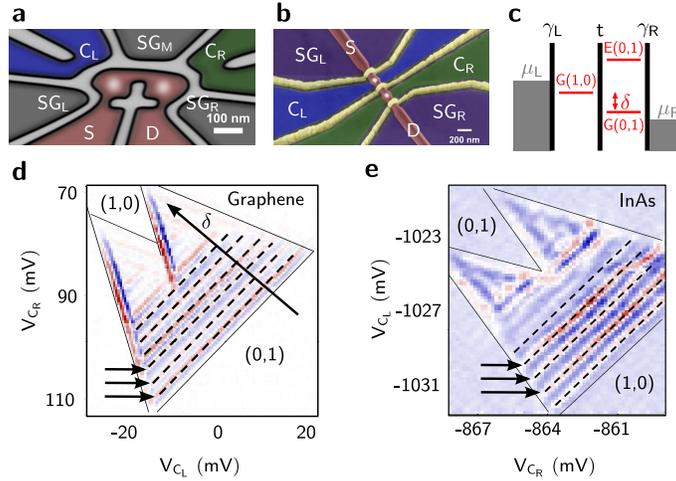}}
\caption{\textbf{Current in graphene and NW DQDs.} \textbf{(a)}
Schematic representation of the graphene DQD. The two dots are
separated by a 30 nm wide constriction and connected to source and
drain (in red) by 20 nm wide constrictions. Both constrictions, in
blue and green, serve as side gates to control the electrochemical
potential of the QDs as well as charge detectors for the QDs.
Additional side gates are shaded in gray. \textbf{(b)} Tilted
scanning electron microscopy image of the NW DQD. The InAs NW (in
red) is deposited on an AlGaAs/GaAs heterostructure with a
two-dimensional electron gas 37 nm below the surface. Again, the
constrictions in blue and green serve as side gates and as charge
detectors. The purple side gates offer additional tunability.
Metallic top gates (in yellow) enable us to independently tune the
tunnel barriers. \textbf{(c)} Energy level diagram of the DQD at
finite bias for non-zero detuning.
 \textbf{(d)} Graphene DQD: close-up of one pair of triple
points for $V_{\textrm{bias}}$~=~6 mV at $T=120$ mK (axis $V_{C_R}$
is upside-down). The numerical derivative of the current with
respect to $V_{C_R}$ and $V_{C_L}$ is plotted to highlight the
periodic lines parallel to the base line. \textbf{(e)} InAs DQD:
close-up of one pair of triple points for $V_{\textrm{bias}}$~=~2.5
mV at 130~mK. The numerical derivative of the current with respect
to $V_{C_R}$ and $V_{C_L}$ is plotted. }\label{figure1}
\end{figure}

\textbf{Results}\\
\textbf{Observation of periodic oscillations}. The two investigated
devices are shown in Fig.~1a (graphene) and Fig.~1b (InAs-NW). The
current through a DQD is maximal at triple points of the charge
stability diagram, where the electrochemical potentials of both dots
are degenerate and aligned with the electrochemical potentials of
source and drain \cite{Ha07RMP79p1217}. The schematic configuration
of the DQD is illustrated by the energy level diagram in Fig.~1c. In
Figs.~1d and e, we show the numerical derivative of the current with
respect to $V_{C_R}$ and $V_{C_L}$, $\partial^2 I/\partial
V_{C_R}\partial V_{C_L}$, for one pair of triple points in each
material system. A bias voltage $V_{\textrm{bias}} = 6$ mV
(graphene) and 2.5~mV (NW) has been applied across the DQDs which
results in a triangular shaped region of allowed transport.

Along the baseline of the triangles (in Figs.~1d and e), the two
ground-state levels G(1,0) and G(0,1) are aligned and $\delta=0$. We
can roughly estimate the number of electrons in each InAs NW dot to
N~$\sim$~30. In graphene, a similar estimation is very difficult
since we do not know exactly where the Dirac point is located. For a
detuning $\delta\neq0$, \textit{inelastic transitions} are probed:
if energy can be exchanged with the environment, a current flows as
observed in Figs.~1d,e. A striking feature is the presence of
periodic peaks parallel to the baseline, indicated by arrows in
Figs.~1d,e, with a periodicity $\delta_0=430~\mu\textrm{eV}$
(graphene, first cool-down) and a smaller periodicity
$\delta_0=190~\mu\textrm{eV}$ (NW). These periodic current
modulations have been observed in different triangles, in both bias
directions with equal periodicity, and in two different cool-downs
(for graphene).

For a more detailed analysis of similarities and differences between
graphene and InAs NW DQDs, the currents through the DQDs along the
detuning line (indicated in Fig.~1d) have been measured (Fig. 2) in
a different triangle. As in the previous measurement, we observe
that the periodicity in graphene is larger (270 $\mu$eV) than in the
InAs NW (200 $\mu$eV). The periodic peaks are more pronounced in the
InAs NW than in graphene. In total we have measured 4 pairs of
triple points for different gate configurations in graphene and have
extracted $\delta_0\approx 430~\mu\textrm{eV}$ for the first
cool-down and $\delta_0\approx 220-280~\mu\textrm{eV}$ for the
second cool-down. In the InAs NW we have measured 19 different
triple points, 15 being in a range $\delta_0\approx
150-215~\mu\textrm{eV}$. These values are much smaller than the
typical excited state energies which are several meV above the
ground state (see Figs.~2a and b), due to the small effective mass
$m^{\star}\approx~0.02~m_0$
\cite{Web10PRL104p036801,Pfund09PRB76p161308}. In graphene,
measurements realized a in single dot of similar size show
excitations at energies of around 2-4 meV
\cite{Sc09APL94p012107,Gut09PRL103p046810}. In the measured device
we see current steps superimposed on the periodic current
oscillations at energies of ~2 meV. We attribute the stepwise
current increase to the opening of additional transport channels due
to excited states, with comparable energies as measured in single
dots. In the following, we provide evidence that our observations
are related to an interference effect.

\begin{figure}
\centerline{\includegraphics[angle=0,width=9cm,keepaspectratio,clip]{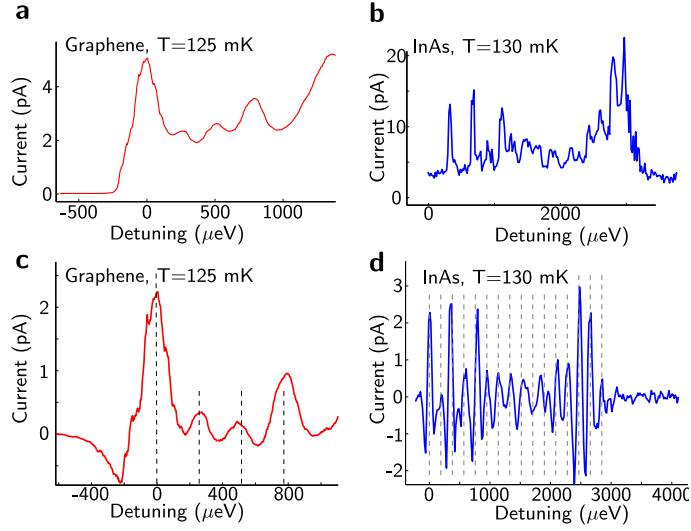}}
\caption{\textbf{Periodic oscillations in the current.} \textbf{(a)}
Raw experimental data: current through the graphene DQD along the
detuning line for a bias voltage $V_{\textrm{bias}}$~=~-5 mV at
$T=125$ mK. \textbf{(b)} Raw experimental data: current through the
InAs NW DQD along the detuning line for a bias voltage
$V_{\textrm{bias}}$~=~3 mV at $T=130$ mK. \textbf{(c)} Current
through the graphene DQD along the detuning line (smoothed and
background subtracted with an adjacent averaging algorithm). A
periodicity of about 270 $\mu$eV can be observed. \textbf{(d)}
Current through the InAs DQD along the detuning line (smoothed and
background subtracted). We observe pronounced periodic oscillations
with an amplitude of few pA. }{\label{figure2}}
\end{figure}

\textbf{Features of the coherent coupling}. To understand these
observations, we model inelastic transitions which occur in a DQD.
We first suppose that for inelastic transitions phonons are emitted
\cite{Fuji98Science282p932}. Therefore, the transition rates are
proportional to the Bose-Einstein coefficient. For graphene, the
coupling of acoustic phonons to electrons is due to deformation
potential coupling
\cite{Bolotin08PRL101p096802,Hwang08PRB77p115449,Vasko07PRB76p233404}.

The double dot is modeled as a two level system with a bonding state
$\ket{\Psi_+}=c_L\ket{L}+c_R\ket{R}$ and an antibonding state
$\ket{\Psi_-} =c_R\ket{L}-c_L\ket{R}$ with $|c_L|^2+|c_R|^2 = 1$. We
are interested in transitions between these two states due to the
absorption or emission of acoustic phonons. The relevant matrix
element for this process is $\bra{\Psi_+}e^{i\textbf{qr}}\ket{\Psi_-}$, where
$e^{i\textbf{qr}}$ arises from the Hamiltonian that describes the
electron-phonon coupling. Emission
and absorption rates due to the deformation potential coupling can be written as
\cite{Ross09Spinger}
\begin{equation}
\begin{split}
\Gamma_{abs/em}=\frac{\pi
D^2}{NM}\sum_\textbf{q}\frac{1}{\omega_{\lambda}(\textbf{q})}q^2(n(\textbf{q})+\frac{1}{2}\pm\frac{1}{2})
\\
\times|\bra{\Psi_+}e^{i\textbf{qr}}\ket{\Psi_-}|^2\delta(\Delta-\hbar\omega_\lambda(\textbf{q})),
\end{split}\label{eq1}
\end{equation}
where $M$ is the single atom mass, $N$ the number of atoms in the
system and $\Delta$ the energy splitting between the two levels. In
order to simplify $|\bra{\Psi_+}e^{i\textbf{qr}}\ket{\Psi_-}|^2$, we
make two assumptions. First, we neglect the overlap between the
electronic wave functions of the two dots, which is equivalent to
saying that the cross terms
$(c_R^2\bra{L}e^{i\textbf{qr}}\ket{R}-c_L^2\bra{R}e^{i\textbf{qr}}\ket{L})$
can be neglected compared to the direct terms
$c_Rc_L(\bra{L}e^{i\textbf{qr}}\ket{L}-\bra{R}e^{i\textbf{qr}}\ket{R})$.
This gives $\bra{\Psi_+}e^{i\textbf{qr}}\ket{\Psi_-}=
c_Rc_L(\bra{L}e^{i\textbf{qr}}\ket{L}-\bra{R}e^{i\textbf{qr}}\ket{R})$,
i.e. there are two alternative decay paths. In a second step, we
assume that $\bra{\textbf{r}}L\rangle$=$\phi(\textbf{r})$, where
$\phi(\textbf{r})$ describes a state localized in the left dot which
typically has an exponential decay at the dot boundaries, and
$\bra{\textbf{r}}R\rangle$=$\phi(\textbf{r}-\textbf{d})$ the state
in the right dot, with $\textbf{d}$ being the distance vector
between the two dots. From this we obtain the relation
$\bra{R}e^{i\textbf{qr}}\ket{R}=e^{i\textbf{qd}}\bra{L}e^{i\textbf{qr}}\ket{L}$.
We finally have $\bra{\Psi_+}e^{i\textbf{qr}}\ket{\Psi_-} =
c_Rc_L\bra{L}e^{i\textbf{qr}}\ket{L}(1- e^{i\textbf{qd}})$, and
therefore
$\Gamma_{abs/em}\propto|\bra{\Psi_+}e^{i\textbf{qr}}\ket{\Psi_-}|^2\propto(1-\cos(\textbf{qd}))$.
This expression predicts periodic oscillations in the emission rate
as a function of quasi-momentum  $\textbf{q}$, which can be
interpreted as an interference effect in which the electron relaxes
in either the left or right dot, analogous to the Dicke subradiant
spontaneous decay \cite{Dicke54PR93p99,Brand99PRL83p3021}. The phase
difference between the two interfering paths is related to the
phonon wave function and the opposite parity of ground- and excited
state (see Fig.~3a).

The periodicities in graphene allow to extract a distance
between the two dots of $d=144$ nm (first cool-down) and $d=220-280$ nm
(second cool-down) for a sound velocity of
$v=15\times10^3~\textrm{m/s}$ (25\% below the theoretical
predictions \cite{Hwang08PRB77p115449}) and $d=96-137$ nm for a
sound velocity $v=5\times10^3~\textrm{m/s}$
\cite{Web10PRL104p036801} in the InAs NW DQD. In both
cases, our results are in agreement with the geometry of the DQDs
(in the graphene DQD,  we have measured a mutual coupling energy
$E_{c,1}^m~\approx~2.4~$meV (first cool down) $E_{c,2}^m~\approx~
1.7~$meV (second cool down) in agreement with a larger effective
distance between the two dots for the second cool down).

In a recent pioneering experiment realized in an InAs NW DQD with a
diameter of 50 nm~\cite{Web10PRL104p036801}, similar equidistant
peaks with a separation $\delta\approx 180-200~\mu$eV were observed.
The authors argue that these resonance lines result from a modified
phononic energy spectrum due to the strong radial confinement.
Following this argument we should observe a separation
$\delta\approx 70-75~\mu$eV in our measurements, considering that
our NW is thicker (130 nm). Instead, we measure (at different triple
points and for different gate configurations) $\delta\approx
150-215~\mu$eV. We also find that the periodicity $\delta$ is
strongly affected by the electrostatic potential of our gates (not
shown), although the phonon energy spectrum should not depend on the
electronic environment. Thus we conclude that the interpretation
related to simple phonon confinement \cite{Web10PRL104p036801}
cannot explain our measurements. Nevertheless our interference-based
approach could be in agreement with C. Weber \textit{et al.}'s
observations of a separation of $\delta\approx 180-200~\mu$eV.

\begin{figure}
\centerline{\includegraphics[angle=0,width=9cm,keepaspectratio,clip]{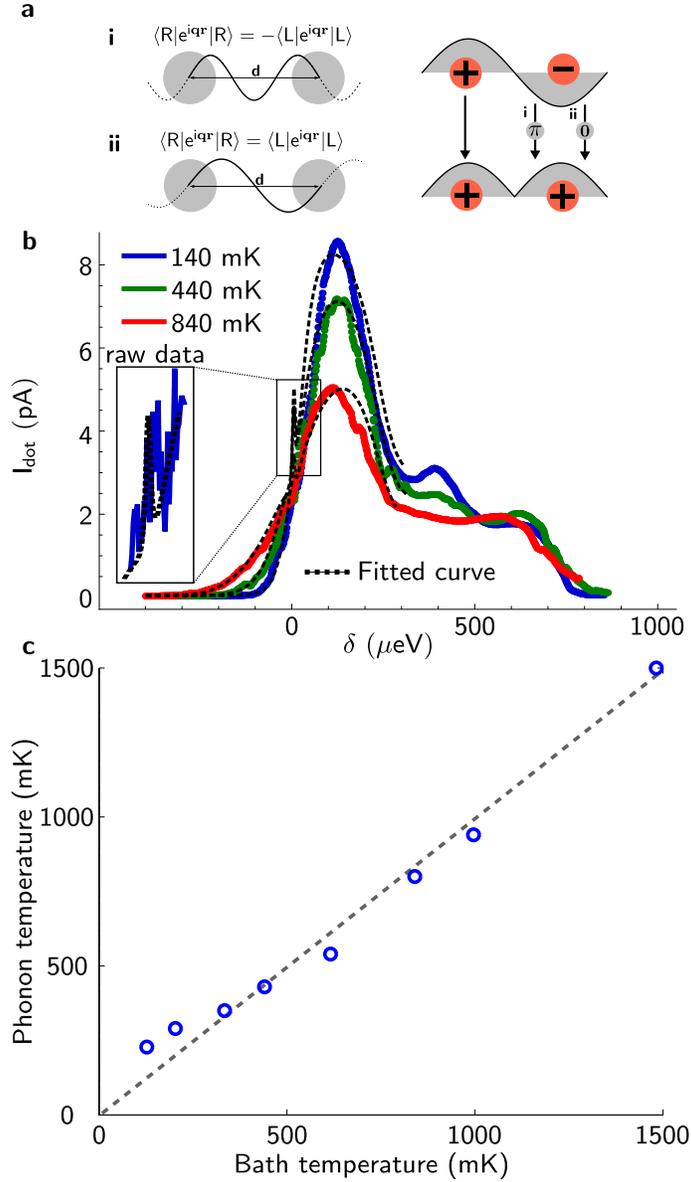}}
\caption{\textbf{Interaction with a bosonic environment.}
\textbf{(a)} Intuitive picture of the interference-based
interpretation of the phonon effect. Depending on whether the
distance between both dots is an even or an odd multiple of half a
phonon wavelength, the phonon mediated transition will shift the
overall phase of the electronic wavefunction by $k\times2\pi$ (case
ii) or $(k+1/2)\times2\pi$ (case i). In the antibonding state,
transitions change the overall phase by $\pi$, in the bonding state
the phase remains unchanged for an electron in the different spatial
eigenstates. Overall this leads to constructive interference in the
right dot for case i (wavefunctions in phase) and destructive
interference for case ii (phase difference of $\pi$). \textbf{(b)}
Current through the graphene DQD as a function of the detuning
between the two dots for $V_{\textrm{bias}}$ = -1 mV. Experimental
data (solid lines): Inelastic transitions close to $\delta$~=~0 are
strongly affected by temperature, as expected if we suppose a
bosonic environment. The overall current is also modified since the
coupling to the leads is slightly changed by temperature. Fitted
curve (dashed lines): our model includes deformation potential
coupling to phonons, coupling to the leads and the phonon absorption
and emission rates. The broadening at $\delta=0$ comes from the
temperature dependence of the phonon absorption and emission rates,
given by the Bose-Einstein distribution. Inset: The main figure
shows averaged measurements (the measurements have been repeated 5
times) which wash out the elastic transition at $\delta=0$. If we
focus on a single measurement (as shown here), we observe a peak at
zero detuning in most measured curves. \textbf{(c)} From the fitted
curves we extract the phonon temperatures plotted as a function of
bath temperature.}{\label{figure3}}
\end{figure}

\textbf{Comparison between InAs nanowire and graphene}. The
measurements of Figs.~2c and d indicate that the interference effect
is more pronounced in the InAs NW than in graphene. Different
mechanisms could lead to such an observation. For instance, the
assumption of localized electronic wavefunctions without overlap
could be less relevant in graphene leading to a damping of the
oscillations. In addition to this, the dimensionality of the system
could influence the amplitude of the interference. In a 2D system
for example, the oscillatory part will be proportional to
$\int_0^{2\pi} d\theta [1-\cos(d\Delta/\hbar c_{\lambda}
\cos(\theta))]$, where $c_{\lambda}$ is the sound velocity and
$\theta$ is the angle between the quasi-momentum $\textbf{q}$ and
the distance vector $\textbf{d}$. In an ideal 1D system we would
obtain a factor $[1-\cos(d\Delta/\hbar c_{\lambda})]$ which implies
more pronounced interference. Theoretical calculations which compare
the phonon spectrum in a graphene and in an InAs NW DQD are needed
to further clarify this point.

\textbf{Signatures of a bosonic environment}. In the solid-state
environment of our DQDs, temperature is expected to antagonize
interference effects by destroying coherence. In the phonon
absorption and emission rates (equation \ref{eq1}), temperature only
appears through the Bose-Einstein distribution with
$\Gamma_{em}\propto(\langle n\rangle + 1)$,
$\Gamma_{abs}\propto\langle n\rangle$, and $\langle n\rangle$=
$1/(e^{\Delta/kT}-1)$ where $\Delta=\sqrt{\delta^2+4t^2}$ is the
energy difference between the bonding and antibonding state, $t$ the
tunnel coupling between the dots, and $\delta$~=~$E_L$-$E_R$ the
energy difference between left and right states
\cite{Oost98Nat395p873}. Measurements of the current through the
graphene DQD close to $\delta\sim 0~\mu$eV for different
temperatures are shown in Fig.~3b. We observe a clear dependence of
the inelastic transitions on temperature, and as expected the
current for negative detuning is enhanced at elevated temperatures
since absorption is more pronounced when the temperature is
increased.

For more quantitative predictions, we calculate the expected current
through the DQD: the rate equations for the occupation of the states
can be expressed as a function of the tunnel coupling, $\Gamma_L$
(left barrier), $\Gamma_R$ (right barrier), $\Gamma_{em}$
(emission), and $\Gamma_{abs}$ (absorption), allowing to calculate
the stationary solution \cite{Gasser09PRB79p035303}. The stationary
current is plotted as a function of detuning $\delta$ in Fig.~3b
(dashed line) for a bath temperature of $\sim$ 140 mK. The good
agreement with the measured curve enables us to extract a phonon
temperature $\sim$ 230 mK. The discrepancy between bath and phonon
temperatures disappears when the mixing chamber temperature is
increased, as can be seen in Fig.~3c, where the extracted phonon
temperatures are plotted as a function of bath temperature. Above
300 mK, bath and phonon temperatures are in excellent agreement.
This result confirms an energy-exchange with the phonon bath.

\begin{figure}
\centerline{\includegraphics[angle=0,width=11cm,keepaspectratio,clip]{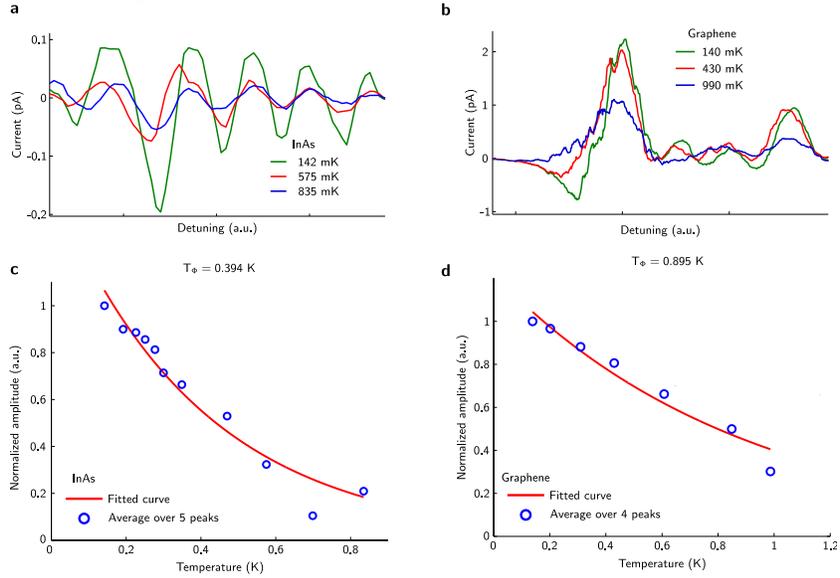}}
\caption{\textbf{Temperature dependence of current oscillations.
}\textbf{(a)} InAs DQD: current oscillations with background
subtracted using an adjacent averaging algorithm, for different
temperatures as a function of detuning. A magnetic field of 1 T is
applied in order to have a better resolution of the peaks.
\textbf{(b)} Graphene DQD: current oscillations with background
subtracted, for different temperatures as a function of detuning.
\textbf{(c)} Normalized amplitude of the oscillations as a function
of temperature for the InAs DQD (quasi-1D system). The blue circles
represent the normalized oscillation amplitude, averaged over the
first five maxima. The red solid line is a $\exp{(-T/T_\phi)}$-fit,
yielding $T_{\varphi}=394$ mK. \textbf{(d)} Normalized amplitude of
the oscillations, averaged over the first four maxima, as a function
of temperature for the graphene DQD (2D system). The average
temperature dependence of the first four peaks is similar and can be
fitted with an exp(-$T/T_{\varphi}$) (red solid line) behavior with
$T_{\varphi}=895$ mK. }{\label{figure4}}
\end{figure}

\textbf{Temperature dependence}. In Figs.~4a and b, we plot the
current oscillations for InAs and graphene DQDs, respectively, for
temperatures from 140 mK up to 1 K. We note a continuous decrease of
the peak amplitudes with increasing temperature. These observations
confirm the difference compared to the interpretation in Ref.~\cite
{Web10PRL104p036801}: current replica due to the interaction between
electrons and individual phonon modes should not depend strongly on
temperature below 1 K ($\delta_0\approx 200~\mu$eV~$\approx~2.3~$K).

The extracted temperature dependence of the average amplitude for
the first 5 peaks (normalized by the lowest temperature amplitude)
is presented in Figs.~4c. The red solid line is a
exp($-T$/$T_\varphi)$-fit to our data with $T_{\varphi}=394$ mK as
discussed below. A similar dependence is observed in graphene
(Fig.~4d) with $T_{\varphi}=895$ mK.

\textbf{Discussion}.\\ To explain this temperature dependence, we
consider decoherence: we suppose that the phase difference between
the two interfering emission processes is not perfectly defined due
to coupling to the thermal bath. In the limit where the thermal
energy is large compared to the energy equivalent of the transition
time scales (\^= $\mu$eV in our case), phase fluctuations are given
by the fluctuation-dissipation theorem and the visibility $\nu$ of
the oscillations decays exponentially over a characteristic
temperature $T_\varphi$: $\nu$ $\propto$
$e^{-\langle\delta\varphi^2\rangle}$ = $e^{-T/T_\varphi}$
\cite{Stern90PRA41p3436}. The quantity $T_\varphi$ quantifies the
strength of the thermal fluctuations the electron is experiencing in
the interferometer (see Fig.~3a). Since decoherence of the phonon
wave function is quite unlikely, we think that with increasing
temperature bonding and antibonding states are washed out. However,
theoretical calculations are needed to validate this hypothesis.

In conclusion, we have observed an oscillatory effect in transport
through DQDs in graphene and in an InAs NW which we attribute to
coherent electron-phonon coupling in nanostructured geometries. The
characteristic phase difference for this interference is determined
by the phonon field phase difference between both dots. We have
observed that the effect is more pronounced in an InAs NW compared
to a graphene DQD. Finally, we have shown that the temperature
dependence of these oscillations is compatible with a thermal
decoherence mechanism. This study may contribute to a better
understanding of the role of the phonon bath in electron transport
through NWs and graphene DQDs.


\begin{methods}
\textbf{Fabrication}. The procedures to fabricate our graphene
quantum system are described in Refs.
\cite{Mo09APL94p222107,Mo10EL89p67005}. The InAs NW, grown by metal
organic vapor-phase epitaxy, is deposited on a predefined Hall bar
of an AlGaAs/GaAs heterostructure with a two-dimensional electron
gas (2DEG) 37~nm below the surface. The etching mask for the DQD and
detector structure, realized by electron beam lithography, is
designed in such a way that the trenches in the 2DEG forming the
detectors and the constrictions in the NW are aligned (see Fig.~1b)
\cite{Sho08NL8p382}. Applying a voltage between each detector and
the corresponding QD enables us to control the electrochemical
potentials of the QDs \cite{Choi09NJP11p013005}. In order to tune
the tunnel barriers independently, the trenches in the 2DEG are
additionally filled with metallic top gates (AlOx/Al/Ti/Au). The two
dots have a lithographic dimension of roughly 100 nm for the
graphene DQD and 130 nm for the InAs NW. The measurements were
performed in two different dilution refrigerators at 120 mK (base
temperature).

\end{methods}



\begin{addendum}
 \item[Acknowledgements] We thank A. Fuhrer, A. Wacker and U. Gasser for helpful discussions.
 \item[Author Contributions] The experiments were conceived and carried out by
 P.R., S.B., T.C., F.M., J.G., T.M. and S.D. The InAs DQD was fabricated by T.C and
 P.R and the graphene DQD realized by F.M.
 The data were analysed by P.R., S.B. and F.M.
 The manuscript was written by S.B. and P.R. The whole project
was supervised by P.R, K.E. and T.I.
 \item[Competing Interests] The authors declare that they have no
competing financial interests.

\end{addendum}


\end{document}